\documentclass[runningheads]{llncs}

\usepackage{graphicx}
\usepackage{hyperref}
\usepackage[all]{hypcap}
\usepackage{cellspace}
\usepackage{xcolor}
\hypersetup{
    colorlinks,
    linkcolor={red!70!black},
    citecolor={green!70!black},
    urlcolor={blue!80!black}
}

\DeclareMathAlphabet{\pazocal}{OMS}{zplm}{m}{n}

\newcolumntype{C}[1]{>{\centering\arraybackslash}S{m{#1}}}
\newcolumntype{L}[1]{>{\raggedright\arraybackslash}S{m{#1}}}
\newcolumntype{N}{@{}m{0pt}@{}}

\usepackage{mathtools}
\usepackage{amsmath}
\usepackage[linesnumbered,ruled,lined]{algorithm2e}

\begin{document}
\title{ Does Differential Privacy Prevent Backdoor Attacks in Practice? }
\titlerunning{DP against Backdoor Attacks}
\author{Fereshteh Razmi \inst{1} \and
Jian Lou \inst{2} \and
Li Xiong \inst{1}}

\authorrunning{F. Razmi et al.}

\institute{Emory University, Atlanta GA 30322, USA  \\ \and
Zhejiang University, Hangzhou, Zhejiang 310027, China \\ 
\email{\{frazmim,lxiong\}@emory.edu}, \\
\email{jian.lou@zju.edu.cn}}

\maketitle              % typeset the header of the contribution
\begin{abstract}
Differential Privacy (DP) was originally developed to protect privacy. However, it has recently been utilized to secure machine learning (ML) models from poisoning attacks, with DP-SGD receiving substantial attention. Nevertheless, a thorough investigation is required to assess the effectiveness of different DP techniques in preventing backdoor attacks in practice. In this paper, we investigate the effectiveness of DP-SGD and, for the first time in literature, examine PATE in the context of backdoor attacks. We also explore the role of different components of DP algorithms in defending against backdoor attacks and will show that PATE is effective against these attacks due to the bagging structure of the teacher models it employs. Our experiments reveal that hyperparameters and the number of backdoors in the training dataset impact the success of DP algorithms. Additionally, we propose Label-DP as a faster and more accurate alternative to DP-SGD and PATE. We conclude that while Label-DP algorithms generally offer weaker privacy protection, accurate hyper-parameter tuning can make them more effective than DP methods in defending against backdoor attacks while maintaining model accuracy.

\keywords{Differential Privacy  \and Backdoor Attack \and Security}
\end{abstract}

\section{Introduction}\label{section1}

Deep neural networks are vulnerable to backdoor attacks. These attacks insert trigger patterns into the image pixels of training data (backdoors) and cause the samples containing the same triggers to be misclassified during inference time \cite{gu2019badnets}. Backdoor attacks resemble conventional poisoning attacks since they both manipulate the training date to serve the attacker's objective \cite{jagielski2018manipulating}. However, in contrast to conventional poisoning attacks, a backdoor attack retains the inference accuracy for benign samples and only activates in the presence of the trigger. Many studies have proposed countermeasures against this powerful attack. The most common approach in these studies is discovering some abnormalities in model statistics or training data \cite{chan_poison_2019,DBLP:conf/aaai/ChenCBLELMS19,wang2019neural,peri2020deep,tran2018spectral}.

A recent promising area of research focuses on differential privacy (DP) \cite{dwork2006calibrating} to build robust models against backdoor and poisoning attacks. This is accomplished by introducing randomness to the model through DP techniques, making it less sensitive to input. There are a few works exploring this area in theory \cite{borgnia2021dp,10.5555/3367471.3367701}. A few others have obtained experimental results by employing DP techniques, particularly DP-SGD (DP-Stochastic Gradient Descent) \cite{borgnia2021strong,DuJS20,hong2020effectiveness,xu-etal-2021-mitigating-data}. However, these studies fall short of a comprehensive investigation. They provide some evidence that  models trained with DP-SGD mitigate poisoning attacks, but they do not explore the power of other state-of-the-art DP models against backdoor attacks.

This paper aims to bridge the theory and practice and provide a comprehensive and in-depth understanding of whether and, more importantly, how various DP models and methods defend against backdoor attacks in practice given the theoretical promise and preliminary evidence in the literature. We study both the standard DP class of algorithms and the Label-DP variant for the first time against backdoor attacks and compare four representative DP and Label-DP algorithms in their defense power. We evaluate their performance empirically on two widely used datasets in the domain of backdoor attacks and differential privacy, namely MNIST and CIFAR-10. To summarize, we make the following contributions:

\begin{enumerate}

\item \textbf{Comparative study of DP approaches against backdoor attacks, including standard DP-SGD approach and the less-studied PATE approach.} Some studies use DP-SGD for training DP models, to defend against poisoning or backdoor attacks. In this work, we explore another well-known DP algorithm, PATE (Private Aggregation of Teacher Ensembles), and test it against backdoor attacks. 

We compare PATE to DP-SGD and show that these classical DP approaches can provide robust models for backdoor attacks. Also, we will demonstrate that the bagging structure of the PATE inherently makes it suitable against backdoors.

\item \textbf{A deeper understanding of the impact of noise and other parameters of DP approaches on backdoor attacks.} The effectiveness of DP approaches is affected by other parameters besides noise. We explore the origin of these algorithms' resilience by examining whether randomness is the sole player or if the other parameters have an impact. 

We empirically show that the randomness (privacy budget) contributes to mitigating the backdoor attack success rate which is compatible with the theoretical results in the literature. However, we demonstrate that the impact of other parameters can be significant on the outcome, especially for PATE, e.g. the threshold utilized to aggregate the teacher models' outputs.

\item \textbf{Comparative study of Label-DP approaches against backdoor attacks.} 
Label-DP protects the privacy of the labels of the training data by ensuring the output model is indistinguishable with respect to the label of a training sample. We study the Label-DP class of algorithms for the first time against backdoor attacks including two algorithms ALIBI \cite{malek2021antipodes} and LP-2ST \cite{ghazi2021deep}, with two incentives. First, based on the definition of label-DP, we expect it to break the tight association between the backdoor triggers and their assigned target class. Second, Label-DP methods usually converge faster than regular DP algorithms with a higher model utility. It is because the indistinguishability is only required on the labels, hence less noise is required to achieve the same level of privacy.

Our evaluations confirm that Label-DP makes the model more immune to backdoor attacks while maintaining model accuracy. We show that Label-DP is superior to DP approaches in terms of convergence speed. Furthermore, we demonstrate they can achieve better robustness accuracy trade-offs under certain settings. For instance, for a lower percentage of backdoors, ALIBI can eradicate the negative impact of the attack while achieving the highest accuracy among all the other approaches. For stronger attacks with higher backdoors, LP-2ST outperforms other approaches when the privacy budget is low. 

\end{enumerate}

\section{Preliminaries}

\subsection{Backdoor Attacks} \label{sec:backdoorattacks}
Backdoor attacks are a category of attacks that involve attaching a small patch to a portion of a base class of the training dataset along with flipping their labels to a specified target class. After the model has been trained using these backdoor samples, it would be vulnerable to the presence of the patch in the inputs. So as the next step of the attack, the attacker attaches the same patch to some desired test samples of the base class and passes it to the backdoored model, so that this combination of base class pattern plus the patch pattern mislead the model to misclassify the sample as the target class. This form of backdoor attacks initially introduced by Gu et al. \cite{gu2019badnets} are powerful attacks and have gained much attention. Some other works tried to make some other type of backdoor attacks that are less detectable or employ them in other domains including videos \cite{saha2020hidden,zhao2020clean}.  

\subsection{Differential Privacy and Label Differential Privacy}
\subsubsection{Differential Privacy (DP). } DP is a privacy-preserving notion that makes an observer unable to tell if particular information contributes to the outcome \cite{dwork2006our}. In the context of machine learning, a DP method should not reveal whether a training sample has been utilized in the training process. 

Let $X$ and $Y$ be the feature and label domain, respectively. Also, let the training dataset consists of $n$ samples from a domain $U=(X\times Y)_n$. Given sample $x$, we have a classification task for the model M to predict $y$. A randomized training algorithm $\mathcal{M} : U\rightarrow R$ is $(\varepsilon, \delta)$-DP if for any two adjacent datasets $D,{D}'\in U$ differing on at most one sample, it holds that:
\begin{equation}
    \forall S\subset R, P[M(D)\in S] \leq e^{\varepsilon}P[M({D}')\in S] + \delta.
    \label{eq:DP}
\end{equation}

A smaller $\varepsilon$ guarantees stronger privacy but typically leads to a lower utility or accuracy of the model due to the randomization in the training. Using a DP property called \textbf{group privacy}, this definition can be extended to two datasets differing in $k$ examples where $k$ denotes more than one data point \cite{664388}. It is achievable by a linear increase in the privacy cost.

% On the other hand, label differential privacy (Label-DP) considers the labels as the only sensitive part of the training data that requires to be kept secret. 

\subsubsection{Label Differential Privacy (Label-DP). } Label DP is an extension of DP that considers the labels as the only sensitive part of the training data that requires to be kept secret. So in contrast to $(\varepsilon, \delta)$-DP that defines privacy for $D$ and ${D}'$ differing on at most one sample, $(\varepsilon, \delta)$-Label-DP considers $D$ and ${D}'$ differing on \textbf{the label} of at most one sample. Therefore, Label-DP can be seen as a relaxation of DP algorithms that guarantees only the privacy of the labels. One of the applications of Label-DP is recommendation systems where the user's profile or search queries are public, but the history of the user rating is sensitive.

\subsection{DP and Label-DP Algorithms for Deep Learning} 
\label{sec:allAlgorithms}
In this section, we explore the main methods for achieving DP (DP-SGD, PATE) and Label-DP (LP-MST and ALIBI) respectively, with Table~\ref{tab:params} showing the critical parameters of the two first algorithms.

\subsubsection{DP-SGD} \cite{abadi2016deep} is the most widely used algorithm for building DP models. DP-SGD restricts the privacy loss in each iteration of SGD (Stochastic Gradient Descent), by updating model in two steps: 1) clipping the L2 norm of the gradients, and 2) inserting calibrated Gaussian noise into those clipped gradients.
% 1. Clipping the L2 norm of the gradients: It bounds the information about input training data in the process of gradient update. This value cannot guarantee differential privacy, but it controls the gradients' sensitivity to the noise. 
% 2. Inserting noise into the clipped gradients. This step adds randomness to the model and provides differential privacy. 

% Thus, two key components of DP-SGD are: 
% \begin{enumerate}
% \item The standard deviation of the noise ($\sigma$).
% \item The upper bound of the clipping norm ($Cnorm$). 
% \end{enumerate}\\

\subsubsection{PATE} \cite{papernot2016semi} provides privacy through a teacher-student structure. First, an ensemble of teachers is trained on disjoint subsets of the private data. Then, given an unlabeled public dataset, a student model queries the teacher ensemble and uses their noisy aggregated vote as the label. The number of queries is restricted. Plus, their response is based on a noisy aggregation without access to any specific private data point. %All leads to guaranteed privacy. 
However, access to a public dataset forces a strong assumption on PATE compared to DP-SGD.

PATE was originally introduced with Laplacian noise \cite{papernot2016semi}. Then it was revised to improve the utility and  privacy trade-off through a more confident aggregated teacher consensus, called Confident-GNMax \cite{papernot2018scalable}. In this paper, we adopt the Confident-GNMax version of the PATE framework, which is based on Gaussian noise. 
% The Confident-GNMax PATE is established on three main components:
% \begin{enumerate}
% \item Threshold $T$: Only those queries are selected that more than a certain number of teachers agree on them.
% \item Selection noise with variance $\sigma_1$: To enforce privacy to the previous step, a Gaussian selection noise is added to the aggregator votes before they are compared with the threshold $T$. 
% \item Result noise with variance $\sigma_2$: Adding another noise to the queries selected from the previous step guarantees differential privacy.
% \end{enumerate}

% There are two other hyperparameters that implicitly affect the accuracy and privacy budget of the PATE:
% \begin{enumerate}
% \setcounter{enumi}{3}
% \item Number of teacher models.
% \item Number of queries. 
% \end{enumerate}

\begin{table}[t!]
\centering
\caption{\textbf{Parameters of the DP algorithms}}
\label{tab:params}
\resizebox{\textwidth}{!}{%
% \begin{tabular}{p{1.5cm}|p{11cm}}
% % \toprule
% %  \hline
% %  \multicolumn{11}{|c|}{Country List} \\
% \hline
% Method &  Parameters  \\
% \hline\hline
% % \midrule

% \begin{tabular}{|>{\raggedright\arraybackslash} m{1.5cm} | m{11cm}|}\hline
% \multicolumn{1}{c}{\cellcolor{}\textbf{Method}} & 
%   \multicolumn{1}{c}{\cellcolor{}\textbf{Parameters}}\\\hline

\begin{tabular}{>{\raggedright\arraybackslash}m{15mm}|m{110mm}}
\hline
\multicolumn{1}{|>{\centering\arraybackslash}m{15mm}|}{\textbf{Method}} 
    & \multicolumn{1}{>{\centering\arraybackslash}m{110mm}|}{\textbf{Parameters}}\\
\hline\hline
   \textbf{DP-SGD}& 
    \begin{enumerate}
        \item \textbf{ Noise multiplier} : Added randomness to the model's clipped gradients to provide DP  
        \item \textbf{ Upper bound of the clipping norm ($Cnorm$)} : Bound to clip the L2-norm of the gradients to control their sensitivity to the noise
    \end{enumerate}
\\ \hline
    \textbf{PATE} & 
    \begin{enumerate}
        \item \textbf{Threshold $T$} : Queries exceeding this minimum teachers' aggregation are selected for training the student model
        \item \textbf{Selection noise with variance $\sigma_1$} : Gaussian noise added to the aggregator's votes before applying threshold to enforce privacy
        \item \textbf{Result noise with variance $\sigma_2$} : Noise added to the selected queries after applying threshold to guarantee DP 
        \item \textbf{Number of teacher models} 
        \item \textbf{Number of queries} 
    \end{enumerate}
\\ \hline
\end{tabular}}
\end{table}

\subsubsection{Label Private Multi-Stage Training (LP-MST)} \cite{ghazi2021deep} is a recent work that achieves Label-DP for deep learning. It leverages a modified version of the Randomized Response (RR) algorithm to add noise to the labels \cite{warner1965randomized}.
RR outputs the actual class of a sample or randomly replaces it with one of the other classes. However, the randomness deteriorates the utility. 

Ghazi et al. \cite{ghazi2021deep} alter the RR algorithm to compensate for the utility, by iteratively training the model on disjoint subsets of the dataset.
Then they use the trained model from the previous stage to get the top-K predictions and limit the RR algorithm to those predictions.  
Similar to the main paper, we report our results on LP-2ST with two training stages. 
% There are three critical hyperparameters in training LP-2ST:
% \begin{enumerate}
% \item The portion we split the training dataset between two stages: Various data splits affect the model's utility. More data in the first stage helps the trained model from the first stage provide better priors for the second stage. However, a small number of data points in the second stage may cause the model in the second stage to underfit.
% \item The temperature: It is applied to the logits utilized to calculate the priors as :
% \begin{center}$p_{i}=\frac{exp(z_{i}/T)}{\sum_{j}^{}exp(z_{j}/T)}$ \end{center}
% where $p_i$ and $z_i$ are the prior and logit of the class $i$ on an input, and $T$ is the temperature. A temperature closer to zero boosts the confidence of the top classes, while a greater temperature makes the priors more uniform. 
% \item The epsilon $\varepsilon$: It is equivalent to the privacy budget of the Label-DP algorithm and induces randomness in RR algorithm. 
% \end{enumerate}

\subsubsection{Additive Laplace Noise Coupled with Bayesian Inference (ALIBI)} \cite{malek2021antipodes} is another Label-DP method in ML that has been recently proposed.
It first adds a Laplacian noise to one-hot labels, then uses these soft new labels to train the model while preserving Label-DP. Since the post-processing does not affect  differential privacy, Bayesian post-processing de-noises the soft labels iteratively during each step of SGD. The combination of additive Laplacian noise and iterative Bayesian inference increases the utility. 
% 
% The only hyperparameter that concerns ALIBI algorithm is:
% \begin{itemize}
% \item 
% The noise of soft training labels: This noise is Laplacian and is only applied once, so $\delta=0$ and the privacy budget do not depend on the learning parameters.
% \end{itemize}

\section{Related Work} 

DP has recently been highlighted for providing robust models to alleviate the negative impact of poisoning attacks. The rationale is that according to the definition of DP and group privacy, DP models are less sensitive to the impact of one or a group of poisoned data. 
In this section, we go through the literature to investigate where and how differentially private approaches used to defend against backdoor and poisoning attacks. We then find the gaps in the literature, formulate those as research questions, and try to answer them and assess the results empirically. 

There are two lines of work in the literature that considered the defensive power of DP methods on poisoning attacks; theoretical and practical studies.

Ma et al. \cite{10.5555/3367471.3367701} theoretically prove the robustness of DP models and provide a theoretical bound. They assume a training dataset $D$ and an attacker with full knowledge creates some poisoned dataset $\tilde{D}$ from $D$. The poisoned model $\theta_{\tilde{D},b}$ is parameterized through the poisoned data $\tilde{D}$ and noise parameter $b$ of the DP model. The attacker's objective loss $C: \Theta \rightarrow R$ aims to misclassify some targets or disrupts the overall classifier's functionality. Assuming the attacker does not know the exact realization of the noise, then the attack is reduced to :
\begin{equation}
\label{eq:attack_loss}
\min_{\tilde{D}} \quad  J(\tilde{D}) = E_{b}\begin{bmatrix}
C(\theta_{\tilde{D},b})
\end{bmatrix}
\end{equation}

Given k poisoned data, the authors utilize the property of differential privacy in Equation (\ref{eq:DP}) and conclude:
\begin{equation}
\label{eq:DP_bound}
J(\tilde{D}) \geq e^{-sign(C).k\varepsilon}J(D)
\end{equation}

According to Equation (\ref{eq:DP_bound}) the attacker is unable to change $J(\tilde{D})$ arbitrarily because it is lower bounded by 0 if $C$ is positive (for example, in case of Mean Squared Error) or it is unbounded from below if $C$ is negative.

This paper provides insight into how DP methods may provide a natural immunity against data poisoning attacks. However, it has two limitations. First, the lower bound of $J(\tilde{D})$ is loose. Second, this paper implements and evaluates its theoretical findings on general attack loss functions and DP frameworks. Thus the specific impact of Equation (\ref{eq:DP_bound}) on SOTA deep learning models (e.g. DP-SGD) and practical attacks (e.g. backdoor attacks) remains neglected. 

To overcome the second limitation, a parallel set of works have employed DP-SGD as a practical usage of DP in deep learning to achieve protection against poisoning attacks \cite{DuJS20,xu-etal-2021-mitigating-data,borgnia2021strong}. Hong et al. \cite{hong2020effectiveness} was one of the first works that considered DP-SGD against backdoor and other poisoning attacks. However, their primary motive was not originated from the fact that DP-SGD is a private algorithm and Equation (\ref{eq:DP}). Instead, they observed that during the training on a poisoned dataset, the gradients computed on poisoned samples have a higher magnitude and different orientation than those computed on clean samples. Hence they leveraged DP-SGD to offset the behavior of the model's gradients on both clean and poisoned data through the randomness of the gradients. 
Their results show some degree of protection against specific poisoning attacks, but their outcome is not promising on backdoor (insertion) attacks. Later, Jagielski and Oprea claimed that DP itself can not serve as a defense against poisoning attacks \cite{jagielski2021does}. They argued that it is possible that the robustness of DP-SGD stems from some parameters other than noise. 

\section{Research Questions}

The existing studies on DP-SGD are inconclusive, and there are no studies on other state-of-the-art DP approaches as a potential defense. It motivates us to extend current works by conducting more comprehensive experiments on DP-SGD and introducing other DP methods as a defense. Based on this primary motivation, we pose some research questions in this section and elaborate their significance. Then in the following sections, we will try to address them empirically. \\

\noindent \textit{\textbf{Question 1.} Is DP-SGD a successful protective algorithm against backdoor attacks? Can PATE, as another main DP approach, mitigate backdoor attacks?} 

\noindent Current studies have differing views on whether DP, particularly DP-SGD, can defend against backdoor attacks. It opens the door for a more comprehensive study of DP-SGD.
% and the impact of its parameters on the backdoor attack.
% On the other hand, the only practical DP method that has been investigated against backdoors is DP-SGD. If that is the only case that DP can protect against poisoned deep learners, then DP models are less likely to be suitable for that purpose. 
It's not clear whether the robustness is achieved by the randomization by DP methods in general or other algorithmic specific parameters of DP-SGD.
Additionally, this outcome can emphasize the gap between DP's theoretical and practical results against poisoning data. 

% So in this work, we first explore DP-SGD to understand why there is no consensus in the literature on DP-SGD as a defensive algorithm. Then for the first time, we explore PATE as a DP method against backdoor attacks to demonstrate if it confirms DP models' robustness. We first explore the effectiveness of these algorithms through their hyper-parameters, and then provide a comparison them directly along with other Label-DP algorithms. The reason that we investigte their heyperparatmeres, even those that are not providing randomness for the differential privacy, is that  Imagine that our experiments suggest that the main reason for PATE defense power originates from a parameter other than noise; we cannot conclude that PATE is a good defense algorithm necessarily because it is differentially private. \\
So in this work, we first explore DP-SGD to understand why there is no consensus in the literature on DP-SGD as a defensive algorithm. Then for the first time, we explore PATE as a DP method against backdoor attacks to demonstrate if it confirms DP models' robustness. We examine the effectiveness of these algorithms by analyzing their hyperparameters, even those that do not contribute to the randomness for DP.
% because it is possible that the primary source of PATE's defense power may originate from a non-random parameter.
With this investigation, we hope to determine whether these algorithms are effective defense mechanism solely because they are DP. \\

\noindent \textit{\textbf{Question 2.} Can other DP notions, such as Label-DP, also provide robustness and even better accuracy and robustness trade-off?  How do different DP notions and algorithms compare in the trade-off?  }
% How does the accuracy-privacy trade-off affect the defensive power of DP algorithms?

\noindent Answering the research question 1, leads us to two other major challenges with regard to DP-SGD and PATE. The first challenge is their prohibitive training time. Training an ensemble of teachers in PATE is heavily costly. Also, DP-SGD requires computation of per-sample gradient norms, which is extremely slow. The other issue with the DP algorithms is the trade-off between the privacy budget and the utility, which means decreasing the privacy budget (i.e., achieving stronger DP) is accompanied by a drop in models' accuracy. We will show that lower privacy budgets usually lead to a lower attack success rate (ASR), which is necessary to defeat attacks. We call this simultaneous reduction in accuracy and ASR the \textit{Accuracy-ASR trade-off}. We will define the criteria for attack success rate in Section \ref{attack_section}. To address these challenges, we conduct a comparison between Label-DP and other DP algorithms by varying DP budgets and attack strengths. \\

% \noindent \textit{\textbf{Question 3.} How do the hidden parameters of the attack and DP methods impact the results?}

% \noindent For the DP and Label-DP algorithms, some hyper-parameters, other than the noise, are embedded in the private algorithms. Each of them may influence these algorithms' defensive power. It is worthwhile to investigate these parameters. Imagine that our experiments suggest that the main reason for PATE defense power originates from a parameter other than noise; we cannot conclude that PATE is a good defense algorithm necessarily because it is differentially private. 

% Hence, we explore these parameters to evaluate whether the algorithms' randomness or other hidden factors protect against the attacks. We also assess the impact of attack strength through the number of backdoors on the defensive approaches. \\

% \noindent Note that these questions are intertwined. It means answering one question can lead to an insight or answer to other questions. We answer all the questions simultaneously and point them out whenever a related question emerges.

\section{Experimental Setup}
\subsubsection{Datasets and Models}
We evaluate each DP model on two datasets: MNIST \cite{lecun1998gradient} and CIFAR-10 \cite{krizhevsky2009learning}.  We study end-to-end training and fine-tuning since both are common practices in modern machine learning. We use the same CNN architecture as \cite{tfprivacy} with two convolutional layers for MNIST and train it from scratch. Also, for CIFAR-10, as \cite{wang2019enhance} suggests, we use ResNet50 \cite{he2016deep} pretrained on ImageNet as a feature extractor and fine-tune its classification head. 

Corresponding to each DP algorithm's specification, we find an optimizer and a learning rate with a grid search algorithm so that the training process achieves the highest accuracy. In addition, data augmentation reduces the effectiveness of all of the attacks \cite{schwarzschild2021just,koh2018stronger}. It leads to a bias in our results. So we skip the data augmentation in our experiments. More details on the training process can be found in the appendix. 

\subsubsection{Attack and Threat Model} \label{attack_section} 
All the DP models are in white-box settings. The backdoors are made based on the triggers introduced in BadNets \cite{gu2019badnets}. To generate backdoors, we first randomly select two classes as base and target class.
% (Horse and Automobile for CIFAR-10 and digit 7 and 1 for MNIST)
Then, we randomly select half of the samples from the base class, attach a $4\times 4$ trigger patch to their bottom right corner and assign the target class as their labels \cite{borgnia2021strong}. We poison 50\% base class to ensure the number of backdoors is high enough, and sufficient clean samples are left in the base class. Under this condition, the model learns both clean and backdoor data points. 

\subsubsection{Evaluation Metrics}
Attack success rate \textbf{(ASR)} is the metric to evaluate the success of the backdoor attacks. According to the definition of the backdoor attacks in Section \ref{sec:backdoorattacks}, ASR indicates the number of test samples from base class that are patched with the backdoor trigger and misclassified as the target class. Thus, a defense method is considered more successful if it leads to a lower ASR. 

The second defensive purpose is to maintain high {\bf accuracy} for the clean test data. The original accuracy of our CIFAR-10 vanilla model over the clean test data is 91.24\% and the backdoor ASR is 98.1\%. The MNIST model's initial accuracy and ASR are 98.92\% and 100\%, respectively.
%CIFAR-10 model trained on only-clean data the accuracy is 92.2\% and if trained on 50\% backdoored data, the accuracy over the clean test data is 91.24\% and the backdoor success rate is 98.1\%. 

\subsubsection{Experimental Roadmap}
This subsection provides an overview of the experiments in the forthcoming sections. 
In Section \ref{sec:dp_experiments}, we analyze two DP algorithms, DP-SGD and PATE, by assessing the impact of their privacy budget and other hyperparameters on the attack success rate. It helps us clarify the underlying reason for their defensive power. At the same time, we will show their resulting accuracy and attack success rate. 
% Then in Section \ref{sec:labeldp_experiments}, we repeat these experiments for two Label-DP algorithms, ALIBI and LP-2ST, to answer similar questions.
Then, in Section \ref{sec:comparison_experiments}, we compare all the DP and Label-DP algorithms in various circumstances to witness which one is prominent and whether the outcome alters in a different situation. Due to space constraints, we could not include all of our experiments and refer to the appendix for our findings on the exploration of parameters for Label-DP algorithms and the training procedure.

\begin{figure*}[t]
  \centering
    \includegraphics[width=0.75\textwidth]{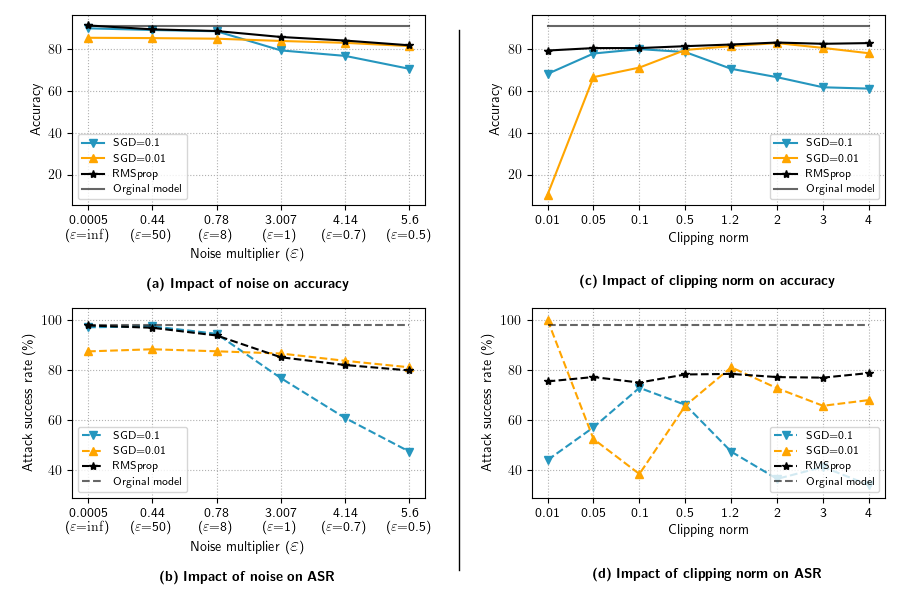}
% \begin{figure}[t]
%   \centering
%     % \includegraphics[width=120mm]{figures/DP-SGD_lr.png}
%     \hspace*{-8mm}
%     \includegraphics[width=90mm]{figures/DP-SGD_lr.png}
    % \vspace{-1em}
    \caption{\textbf{Effectiveness of DP-SGD against backdoor attacks, w.r.t the noise multiplier, clipping norm, and the optimizer.} }
    % The solid lines on the top figures show the accuracy, and the dashed lines show the percentage of attack success rate (ASR). Higher noises can reduce the ASR, but it costs some accuracy reduction as well (Left). Although clipping norms change the accuracy and ASR, this change does not follow a straightforward pattern (Right). For both noise multiplier and clipping norm, the type of optimizer impacts the results significantly.}
    \label{fig:dpsgd}
    % \vspace{-1em}
\end{figure*}

\section{DP against Backdoors} \label{sec:dp_experiments}
This section investigates DP-SGD and PATE, against backdoor attacks. For each algorithm, we will evaluate their key hyperparameters (introduced in Table \ref{tab:params}) on CIFAR-10 dataset and show that some of them have a critical impact on the accuracy and ASR. The results of the MNIST dataset are very similar. So to be concise, we skip their reports here but use them to conduct the experiments in the subsequent sections. 
% LP-2ST, and ALIBI against backdoor attacks. The impacts of the primary hyperparameters introduced in the previous sections are evaluated on the attack on the CIFAR-10 dataset. The results of the MNIST dataset are very similar. So due to the space limitation, we skip their reports here, but use them to conduct the experiments in the subsequent sections.

\subsection{DP-SGD vs. Backdoors}
SGD is the dominant optimizer in practice paired with the DP-SGD algorithm, especially in defeating poisoning attacks \cite{abadi2016deep,borgnia2021dp,hong2020effectiveness,jagielski2021does}. So we consider different optimizers and learning rates to depict the sensitivity of DP-SGD performance to these factors: RMSProp, SGD with a learning rate of 0.1, and SGD with a learning rate of 0.01.
% For CIFAR-10 with $10^{5}$ training samples, a choice of $\delta=10^{-5}$ is suitable . 
Based on the size of the dataset, we set the DP-SGD algorithm as $(\varepsilon,10^{-5})$-DP and report $\varepsilon$ as the privacy budget \cite{papernot2018scalable}. 

Fig. \ref{fig:dpsgd}a and \ref{fig:dpsgd}b show the impact of the noise multiplier by fixing the clipping norm to 1.2 (typical for CIFAR-10).
% The clipping norm is often set between 1 and 2 for CIFAR-10. On the left side of Figure \ref{fig:dpsgd}, we fix the clipping norm to 1.2 and evaluate the impact of the noise multiplier. On the x-axis, the noise multiplier is reported along with the corresponding privacy budget. 
Interestingly, the rate of the accuracy drop to the ASR drop differs for each optimizer. However, in general, the higher noises reduce the accuracy and ASR simultaneously.
% For instance, at noise=5.6, with RMSProp optimizer, the accuracy and the ASR are about 82\% and 80\%. For SGD with a learning rate of 0.1, it is about 70\% and 47\% ,respectively. 
It suggests that SGD can resist the backdoor attacks more significantly by paying slightly more utility cost.

Fig. \ref{fig:dpsgd}c and \ref{fig:dpsgd}d illustrate the impact of different clipping norms on the accuracy (top) and ASR (bottom) using a fixed noise of 5.6. In contrast to RMSProp, for SGD optimizers, the choice of learning rate makes two different patterns of ASR w.r.t the clipping norm, which reveals how SGD training without an adaptive learning rate can be affected by the norm of the gradients. So while the clipping norm significantly impacts the model utility and robustness, it is difficult to optimally adjust it when the defender is agnostic to the attack specifications.

% We fix the noise to 5.6 for Figures \ref{fig:dpsgd}c and \ref{fig:dpsgd}d on the right. These two figures illustrate the impact of different clipping norms on the accuracy (top) and ASR (bottom).
% Depending on the optimizer, the results are very different. Both accuracy and ASR of RMSProp are stable, regardless of the clipping norm. For both SGD optimizers, the accuracy and especially ASR are extremely unpredictable. The choice of learning rate makes two distinct patterns of ASR w.r.t the clipping norm. It reveals how SGD training without an adaptive learning rate can be affected by the norm of the gradients. So while the clipping norm significantly impacts the model utility and robustness, it is difficult to optimally adjust it when the defender is agnostic to the attack specifications. 
According to \cite{bu2022automatic}, the impact of the clipping norm on accuracy is not monotonic, which is manifested as a non-monotonic pattern of accuracy and ASR in Fig. \ref{fig:dpsgd}c and \ref{fig:dpsgd}d. For the reason of the different pattern of ASR in the left side of Fig. \ref{fig:dpsgd}d with SGD-0.01, we speculate that the small learning rate accompanied by a high noise and small clipping norm can hardly learn the normal images' manifold, and instead it retains the repetitive and striking patterns of the backdoor triggers.\\

% Bu. et al. \cite{bu2022automatic} have conducted an ablation study on the joint effect of learning rate and clipping norm and showed that the impact of the clipping norm on accuracy is not monotonic, which is manifested as a non-monotonic pattern of accuracy in Figures \ref{fig:dpsgd}c. Furthermore, the relation between ASR and clipping norm (Figures \ref{fig:dpsgd}d) follows the same pattern, except for the smallest clipping norms on SGD-0.01. For those exceptions, we speculate that an optimizer with a low learning rate accompanied by a high noise and small clipping norm can hardly learn the normal images' manifold. Instead, it retains the repetitive and striking patterns of the backdoor triggers.
\textbf{Conclusion (Q1):} In our evaluations, DP-SGD was successful in mitigating the impact of backdoor attacks. 
However noise multiplier, clipping norm and training parameters determine the extent of this success.  As a result, differences in these parameters contribute to the varying results reported in previous studies on the effectiveness of DP-SGD as a defense mechanism.
% However, not only the noise parameter $\sigma$ but also clipping the gradients determine the extent of this success. Also, DP-SGD is very sensitive to training parameters. Hence in previous works on DP-SGD it is important to consider the training parameters that are utilized in machine learning models to assess its performance on poisoning attacks. All these other components that are involved in the attack success rate are part of the reasons for the different results that are reported on the defense power of DP-SGD method in previous works.

% \end{figure}
\begin{figure}[t]
  \centering
    \includegraphics[width=\textwidth]{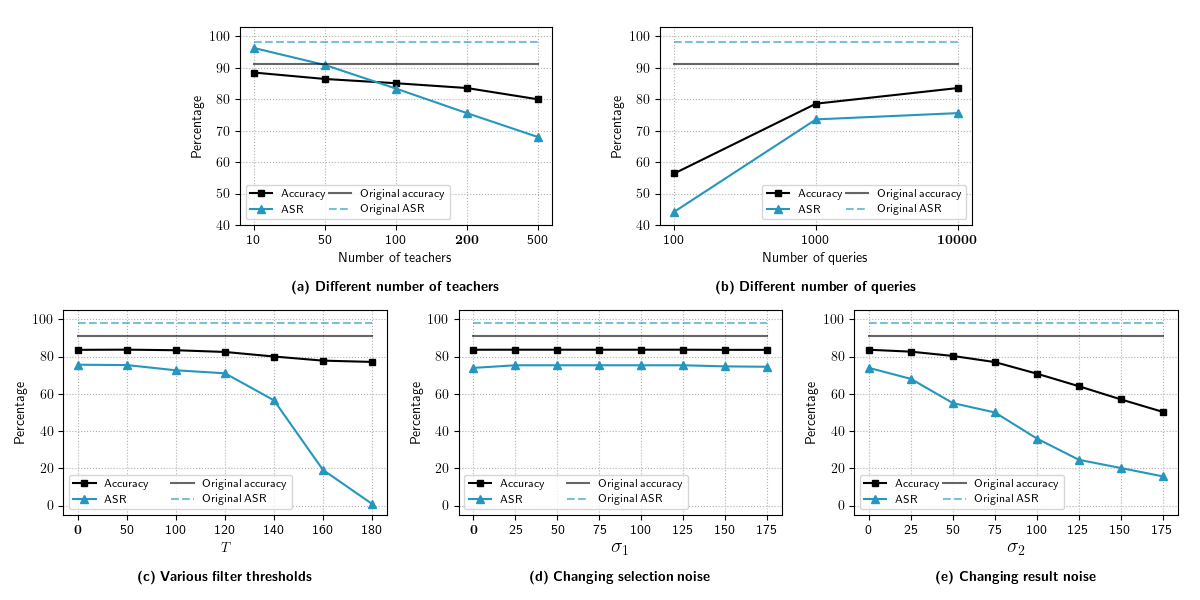}
    % \vspace{-1.5em}
    \caption{\textbf{The impact of number of teachers, number of queries, threshold, selection noise and result noise on the student model's accuracy and ASR from left to right and top to bottom, respectively).}}
    % For figures (a) to (e), we select the optimized value of the discussed metric (highlighted on x-axis) and use it in the next figures. The blue lines represent ASR and the black lines are accuracy. The flat lines are calculated on vanilla model. The number of public queries is bounded to 10000 (b).}
    \label{fig:PATE}
    % \vspace{-1.5em}
\end{figure}

\subsection{PATE vs. Backdoors}
\label{sec:pate}
% PATE consists of teacher models that are combined in the form of bagging. The difference between PATE and other bagging models lies in the three extra components introduced above. Also, PATE use the result of majority voting of the bagging component to train another student model. 
In this section, we evaluate the robustness of PATE against backdoor attacks and the impact of different parameters including number of teachers, number of queries, threshold, selection noise, and result noise. The result is shown in Fig. \ref{fig:PATE}. Whenever noises or threshold are not evaluated, we fix their values equal to 0. In the case of number of queries and number of teachers, the default values are fixed to 10000 and 200, respectively. For training PATE, we assume 1/5 (i.e. 10000 samples) of the training data is publicly available for training the student model, and the rest is private.  In the original PATE paper \cite{papernot2016semi}, the number of queries is set to as low as 1000. However by doing so, we naturally remove a large fraction of poisoned data and make the comparison between different DP methods unfair. So we keep the default number of queries 10000 and in the next sections to compare the models, we analyze the impact of both noise and the number of queries on the PATE's utility and privacy budget. 
% we select the value of the metric the figure is describing and bold it on the x-axis. Then, we will fix that value for the computations in the subsequent figures. 

Fig. \ref{fig:PATE}a and \ref{fig:PATE}b show the number of teachers and the number of queries impact the accuracy and ASR in opposite ways. A higher number of teachers means fewer training data and lower accuracy for each teacher, hence less accurate consensus from the aggregator. It also compromises the consensus on assigning the target class to the backdoor samples and decreases the ASR, which aligns with the literature finding that bagging can hinder the success of the backdoor attacks \cite{biggio2011bagging,jia2021intrinsic,chen2022collective}. Furthermore, in Fig. \ref{fig:PATE}b, lower number of queries are associated with  less training data for the student model and fewer backdoors, hence lower accuracy and ASR.

Fig. \ref{fig:PATE}c illustrates that aggregation threshold is crucial in defeating backdoors and has minimal impact on utility loss.  This finding complements previous results to use bagging against poisoning attacks. The threshold forces the aggregation process to filter out uncertain data and backdoors, resulting in higher accuracy and lower ASR in the student model. To the best of our knowledge, this factor has not been considered in previous works as a major contributor to the effectiveness of bagging.

Fig. \ref{fig:PATE}d and \ref{fig:PATE}e demonstrate the effect of selection noise and result noise used in  selecting and randomizing queries which form the basis of DP for PATE. We found the same trends when one of the noises is fixed to a random positive value. Based on these results, to defeat ASR we need a high result noise which leads to a dropped accuracy. Since we fixed the number of queries and only varied the noise values to control privacy, the privacy budget still remains as large as $\varepsilon=4$ at a high noise level of 175.
% The result noise decreases the ASR but with the cost of high noise and still large privacy budget.
% Although the result noise decreases the ASR but even in the case of a high $noise=175$ the privacy budget is approximately $\varepsilon=4$. The reason that the $\varepsilon$ is still high is because of the large number of queries.
% In the original PATE paper \cite{papernot2016semi}, the query numbers are selected as low as 1000. However by doing so, we naturally remove a large fraction of poisoned data and make the comparison between different DP methods unfair. So here, we keep the query 10000 and in the next sections to compare the models, we analyze the impact of both noise and the number of queries on the PATE's utility and privacy budget. 

\textbf{Conclusion (Q1):} PATE is very successful in defeating backdoor attacks. It can be more successful than DP-SGD but it is highly sensitive to the algorithm parameters. Result noise ($\sigma_2$) and number of queries which are the most influential parameters on the privacy budget ($\varepsilon$) decrease the ASR but they also cause a drastic decrease in the accuracy at the same time. On the contrary, the best result is achieved through tuning the threshold, although it cannot provide any DP by thresholding alone. 

% \begin{figure*}[t]
%   \centering
%     % \includegraphics[width=\linewidth]{figures/trainingProcess.png}
%      \includegraphics[width=\textwidth]{figures/trainingProcess.png}
%         %  \vspace{-1.75em}
%     \caption{\textbf{An overview of the training process of LP-2ST, ALIBI and DP-SGD using $\varepsilon=\infty$ (upper) and $\varepsilon=1$ (lower).}
%     }
%     \label{fig:training}
%         % \vspace{-1.75em}
% \end{figure*}

\section{Comparison of DP and Label-DP Methods} \label{sec:comparison_experiments}
In this section, we compare all the DP and Label-DP algorithms to discover which one and under what conditions are more successful.

\begin{figure*}[t]
  \centering
    \includegraphics[width=\textwidth]{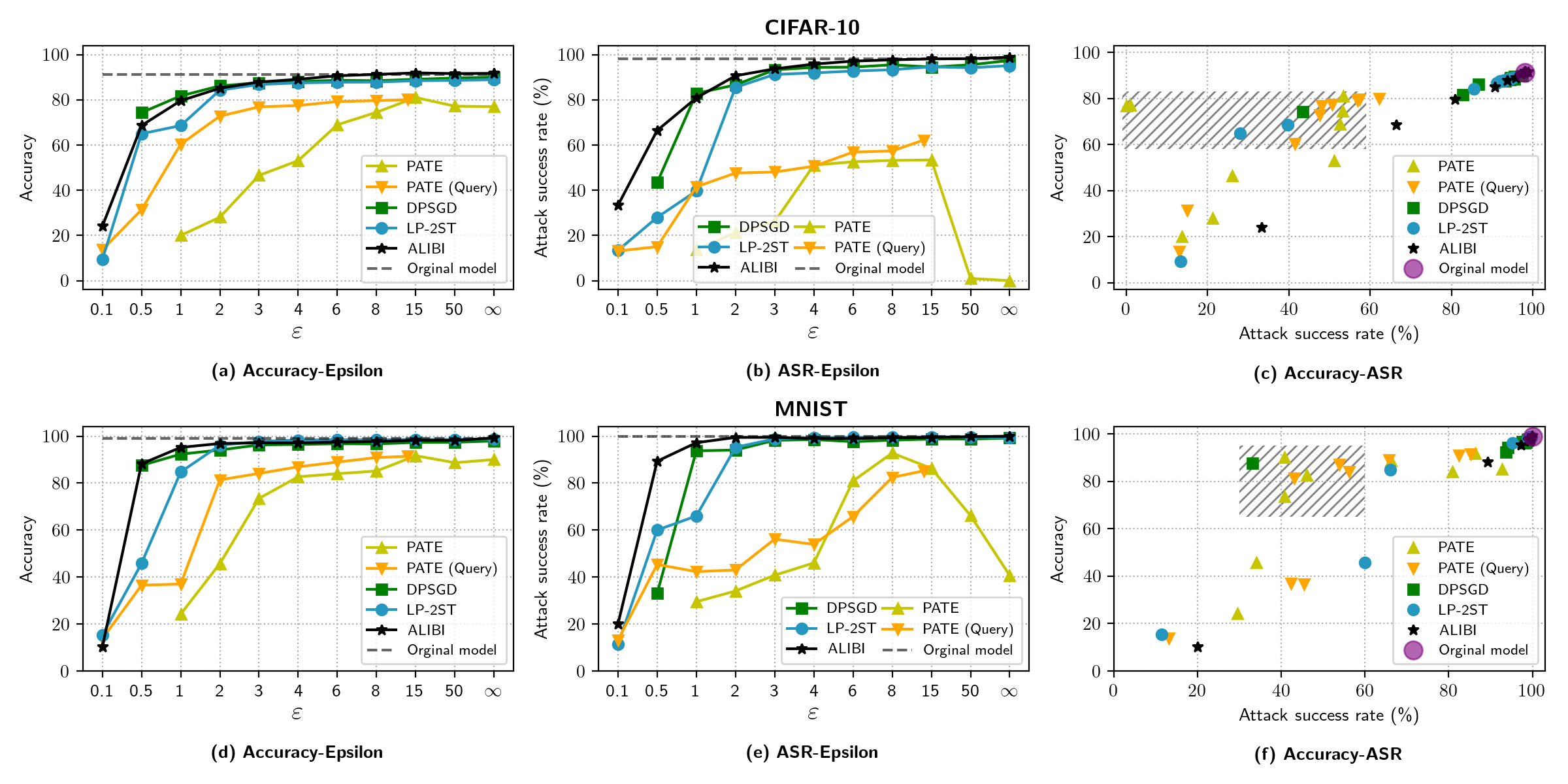}
    % \vspace{-1.75em}
    \caption{\textbf{The impact of epsilon on DP and Label-DP methods using MNIST (top) and CIFAR-10 dataset(bottom).} 
    % For PATE, the impact of changing noises and the number of queries are investigated.
    }
    \label{fig:all_wrt_epsilon}
    % \vspace{-1.25em}
\end{figure*}

\subsection{Privacy Budget Analysis}
The $\epsilon$ in DP and Label-DP serves two different goals. So we do not directly compare the $\epsilon$ values of the two methods even though both can be reduced to label DP \cite{ghazi2021deep}. 
Instead, what we care is the trade-off between accuracy and ASR provided by varying $\epsilon$ of the two methods.  We pick the best parameters from the results in the previous section to conduct the current experiment. The best parameters lead to high accuracy and a low ASR. Wherever there is a trade-off between accuracy and ASR, we prioritize accuracy. For MNIST, we do not present those parameter selections due to the similar outcomes.

% Figure \ref{fig:all_wrt_epsilon}a,b,c compares the accuracy and ASR of the different methods for MNIST. We have repeated all the experiments in the previous section on the MNIST dataset and select the best parameters for the current experiment. However, we skip reporting those parameter selections due to the space limitation and the similar outcome of the CIFAR-10 and MNIST.
% % Therefore, we just use their results to find the best parameters for the MNIST dataset and utilize them in the current experiment (top row of Figure \ref{fig:all_wrt_epsilon}).

Fig. \ref{fig:all_wrt_epsilon}a,b compare the accuracy and ASR of the different methods for CIFAR-10 with varying $\epsilon$ while \ref{fig:all_wrt_epsilon}c shows the trade-off of accuracy and ASR of different methods (the ideal case correspond to 100\%accuracy and 0\% ASR). PATE can achieve different levels of privacy by varying two factors: 1) noises (lime green plots), and 2) number of queries (orange plots).
% Given access to at most 10000 public data points and fixing the noises, we achieved the upper bound of around 12.8 for the privacy budget.
The first observation is that non-DP PATE outperforms all other results and methods (the rightmost point of the lime green plot). It indicates the power of bagging with a threshold against backdoor attacks. LP-2ST for some $\epsilon$ values works well. For instance, $\varepsilon=1$ has high accuracy (78\%) and a significantly decreased ASR (39\%). However DP-SGD gives the best results when $\varepsilon=0.5$. For ALIBI, both accuracy and ASR drop proportionally.

Fig. \ref{fig:all_wrt_epsilon}d,e,f show similar trends for MNIST. Fig. \ref{fig:all_wrt_epsilon}f combines the results of the two other columns by directly comparing the accuracy and corresponding ASR. The rectangular areas with the hatched pattern in the last column consist of the most desired results with high accuracy and dropped ASR regardless of their privacy budget. It includes different private algorithms, but mostly PATE, which indicates the dominance of PATE.

\textbf{Conclusion (Q2):} The DP and Label-DP techniques effectively reduce the vulnerability of backdoor attacks, albeit at the cost of decreased accuracy. If the optimal approach is determined by the accuracy-to-ASR ratio, then the superiority of each DP or Label-DP model depends on the allocated privacy budget.

\begin{figure}[t]
  \centering
    \includegraphics[width=0.8\textwidth]{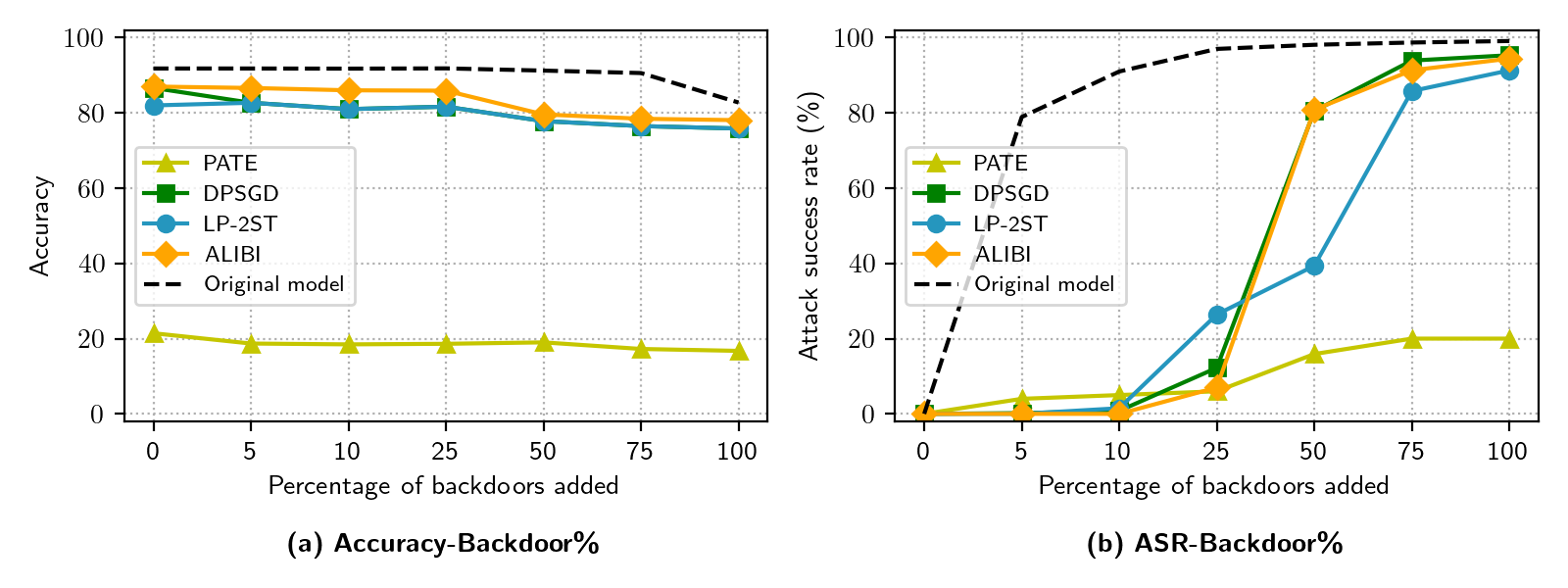}
% \begin{figure}[t]
%   \centering
%     % \includegraphics[width=120mm]{figures/DP-SGD_lr.png}
%     \hspace*{-8mm}
%     \includegraphics[width=85mm]{figures/alleps1_vs_backdoorpercentage.png}
%     % \vspace{-1em}
    \caption{\textbf{The significant impact of poisoned data on DP-based defense methods.} The epsilon is fixed to 1 and then all the methods are compared by varying the percentage of the training data that has been poisoned. 
    % The accuracy does not show a drastic change (Left). However the ASR is very dependent on the number of poisoned data (Right). 
    }
    \label{fig:poison_percent_eps1}
    % \vspace{-1em}
\end{figure}
% \end{figure}

% \vspace{-10}
\subsection{Attack Strength Analysis}
We discussed the hyperparameters and the privacy budget of the algorithm as two factors that impact the immunity of the DP approaches against backdoor attacks. A third factor that should be considered when assessing the level of immunity is the strength of the attack itself. So far, we have synthesized powerful attacks by poisoning 50\% of the data with backdoors. However, in practice, the attacker conceals her malicious activity by limiting the percentage of poisoned data introduced into the pipeline. 
% Therefore scenarios with such high levels of data contamination are unrealistic. 
% In reality, contamination levels are usually lower; as the contamination level is lowered, it becomes harder to uncover the attack. 
Therefore we change the percentage of the backdoors in the base class to develop a range of more realistic and more powerful (but less realistic) attacks. 
% In this analysis, the privacy budget for all the DP algorithms has been fixed to $\varepsilon=1$ to control for that variable.

Fig. \ref{fig:poison_percent_eps1} shows the accuracy and ASR w.r.t number of backdoors, when the privacy budget for all DP algorithms has been fixed to $\varepsilon=1$. We observe that the accuracy does not drastically change w.r.t number of backdoors, yet the ASR increases as the attack becomes more powerful. Looking at the pattern, we can see that the DP algorithms almost entirely diffuse the attack when the percentage of backdoors is sufficiently small. It should be noted that the low accuracy of PATE is a result of controlling its privacy budget by adding noise, rather than limiting the number of queries according to the reasoning we had in section \ref{sec:pate}.

\textbf{Conclusion (Q2):} These results illustrate the effectiveness of DP-SGD, LP-2ST, and ALIBI against more realistic backdoor attacks (with backdoor\% $\leq$ 10). For such attacks, the accuracy drops by 10\%, and the attack achieves no success. This is compatible with Equation~(\ref{eq:DP_bound}) that shows that the attacker's loss limit in DP models is theoretically linked to the number of poisoned data.
% \begin{table}[h!] 
\subsection{Accuracy-Privacy Trade-off}
To see the accuracy when a perfect defense is desired (close to 0 ASR), we have analyzed different privacy budgets for each DP method and found the greatest $\varepsilon$ where the ASR does not exceed 1\%. This small ASR is achievable when the number of backdoors is insignificant (we set it to 10\%). By doing so, we achieve the least randomness that leads to a successful defense. After removing the impact of the attack, we can have a fair comparison of accuracy and training time. 
\begin{table}
\centering
\caption{\textbf{Comparison of the highest accuracy and epsilon that DP methods can achieve while ASR=0.}}
\begin{tabular}{p{2cm}||p{0.2cm}p{2cm}p{2cm}p{2cm}p{2cm}}
% \toprule
%  \hline
%  \multicolumn{11}{|c|}{Country List} \\
\hline
&  &  \textbf{DP-SGD} &  \textbf{PATE}  &  \textbf{ALIBI}  & \textbf{LP-2ST}  \\
\hline\hline
% \midrule
\textbf{Accuracy} &  &  88.67 & 85.02 & \textbf{89.53} & 79.9  \\
\hline
\textbf{Epsilon} &  & 2 & \textbf{inf} & 2 & 0.9 \\
\hline
\textbf{Time} &  & 140s & 220s & \textbf{59s} & \textbf{58s} \\
\hline
% \bottomrule
\end{tabular}
\label{table:acc_privacy}
\end{table}

Table \ref{table:acc_privacy} highlights the best values of accuracy, privacy budget, and training time in each row. 
% LP-2ST shows the weakest accuracy-privacy trade-off, with the lowest accuracy accompanied with a low epsilon. 
The previous findings indicate that a deterministic version of PATE, with noise removed, is the most resilient against attacks. However, when the goal is to simultaneously defend against backdoors and protect privacy, this result is not favorable for PATE. DP-SGD and ALIBI, with the same privacy budget, can achieve better accuracy than PATE.

Finally, with respect to training time, two Label-DP methods demonstrate a considerable reduction in training time, surpassing other DP techniques. It is important to note that this experiment was conducted on a CIFAR-10 fine-tuning task, where training time is negligible. However, in more complex architectures with end-to-end settings, time may become a bottleneck for PATE and DP-SGD.

\textbf{Conclusion (Q2):} 
When a perfect defense is desired, Label-DP methods offers best efficiency and comparable or better accuracy trade-off to DP approaches.

% each DP method has distinct benefits with regards to model utility, training time, or introducing randomness to the model.

% \subsection {The Impact of DP Algorithm Parameters on Backdoors}
% This section investigates the four algorithms DP-SGD, PATE, LP-2ST, and ALIBI against backdoor attacks. The impacts of the primary hyperparameters introduced in the previous sections are evaluated on the attack on the CIFAR-10 dataset. The results of the MNIST dataset are very similar. So due to the space limitation, we skip their reports here, but use them to conduct the experiments in the subsequent sections.

% \vspace{-1em}
% \vspace{-10}
\section{Discussion and Conclusion}

This paper posed important questions regarding the ability of DP to provide robustness against backdoor attacks in practice. In addition to DP-SGD, we explored the other commonly used DP algorithm (PATE) and two Label-DP algorithms (LP-2ST and ALIBI) for the first time for this purpose. We have several main findings. 

First, the noise and randomness added to the private models can indeed decrease the attack success rate of the backdoors, but at the cost of utility drop for clean input. In a nutshell, a model trained with privacy guarantee have inherent benefit in robustness against backdoor attacks. This statement holds for all four methods mentioned above. A somewhat unexpected outcome is that PATE delivers the best results, even without the use of noise (without DP guarantee).

Second, contrary to the claims of some previous studies, DP-SGD provides good resistance against backdoors while keeping the accuracy relatively high. We also observed the same phenomenon for Label-DP algorithms. The accuracy-ASR trade-off is diverse among the DP and Label-DP methods we analyzed. One model may outperform the others depending on the privacy budget, algorithm parameters, and attack specifications. Therefore it is possible to use DP models as defense strategies. A proper selection of the above mentioned factors can adequately balance the accuracy and ASR. 

This work was an empirical study on two benchmark datasets, MNIST and CIFAR-10. %While this paper did not aim to provide a definite theoretical answer to the raised questions, 
It offered new empirical understandings of the connection between DP and backdoor attacks in relation with existing theoretical understandings. Future research could focus on exploring the impact of Label-DP on particular type of poisoning attacks focusing on labels such as label-based flipping attacks. Additionally, given the ability of DP methods to enhance robustness, there is an opportunity to develop modified DP algorithms that offer greater protection against poisoning attacks, and simultaneously fulfill both privacy and robustness objectives.

% \usepackage{graphicx}
% \usepackage{hyperref}
% \usepackage[all]{hypcap}
% \usepackage{cellspace}
% \usepackage{xcolor}
% \hypersetup{
%     colorlinks,
%     linkcolor={red!70!black},
%     citecolor={green!70!black},
%     urlcolor={blue!80!black}
% }
% % \setlength{\cellspacetoplimit}{3.5ex}
% % \setlength{\cellspacebottomlimit}{2ex}
    
% \DeclareMathAlphabet{\pazocal}{OMS}{zplm}{m}{n}
% \newcommand{\La}{\pazocal{L}}
% \newcommand{\Ma}{\pazocal{M}}
% \newcommand{\Xa}{\pazocal{X}}
% \newcommand{\Ya}{\pazocal{Y}}
% \newcommand{\Za}{\pazocal{Z}}
% \newcommand{\Fa}{\pazocal{F}}
% \newcommand{\Ha}{\pazocal{H}}
% \newcolumntype{C}[1]{>{\centering\arraybackslash}S{m{#1}}}
% \newcolumntype{L}[1]{>{\raggedright\arraybackslash}S{m{#1}}}
% \newcolumntype{N}{@{}m{0pt}@{}}

% \usepackage{mathtools}
% \usepackage{amsmath}
% \usepackage[linesnumbered,ruled,lined]{algorithm2e}

\bibliographystyle{splncs04}
\bibliography{ref}
\newpage

\appendix
\section{Appendix}

\subsection{Experimental Setup Details}

\subsubsection{Training Configuration.}
Table \ref{tab:structure} elaborates the details of the architecture of neural networks utilized for the MNIST and CIFAR-10 datasets.
\begin{table}[ht!]
\caption{Model structures used for MNIST and CIFAR-10.}
\footnotesize
\centering
\begin{tabular}{p{3cm} l}
\hline
\tabularnewline
\strut & $Conv(8\times8, 16) \rightarrow ReLU \rightarrow MaxPool $ \\
\textbf{MNIST} ::& $\rightarrow Conv(4\times4, 32) \rightarrow ReLU \rightarrow$ \\
& $MaxPool \rightarrow Linear(32) \rightarrow Linear(10)$\\
\tabularnewline
\hline
\tabularnewline
\textbf{CIFAR-10} ::& $ResNet50 \rightarrow AvgPool \rightarrow$ \\
\strut &  $Linear(256) \rightarrow ReLU \rightarrow Linear(10$)\\
\tabularnewline
\hline
\end{tabular}
\label{tab:structure}
\end{table}

For each DP algorithm, we use a different training configuration so that the training process delivers the model with the highest accuracy. Table \ref{tab:trainSetup1} lists the training setup for each DP algorithm on CIFAR-10. For MNIST the setup had just some minor differences so we did not list them here.
% , including the number of epochs which was 30 for all the DP algorithms. 
For ALIBI in CIFAR-10 experiments, we have a learning rate that is scheduled according to the piecewise constant with linear ramp-up scheme, previously used by \cite{ghazi2021deep}. It increases from 0.003 to 0.1 in the first 30 epochs and then remains piecewise constant at 0.01 and 0.001 in epochs 30 and 40, respectively. We noticed increasing the number of epochs beyond 50 while using various learning rates did not enhance the outcome.

\begin{table}[h!]
\small \caption{Training configurations of four DP algorithms for CIFAR-10. The physical batch size of DP-SGD is set to 128.} \label{tab:trainSetup1}
\centering
\footnotesize
\begin{tabular}{p{2.2cm}|c c c c c}
% \begin {tabular} {m{3pt} | m{4pt}}
\hline
\multicolumn{1}{c}{} & 
  \multicolumn{1}{c}{\textbf{DP-SGD}} &
  \multicolumn{1}{c}{\textbf{PATE}} &
  \multicolumn{1}{c}{\textbf{LP-2ST}} &
  \multicolumn{1}{c}{\textbf{ALIBI}} \\
%   \hline
% Dataset & \\
\hline
\textbf{Optimizer} & RMSProp & Adam & SGD & SGD \\
\textbf{Learning rate} & 0.001 & 0.001 & 0.001 & 0.003 \\% with linear rampup \\
\textbf{Epochs} & 50 & 50 & 50 & 50 \\
\textbf{Batch size} & 512$^*$ & 32 & 128 & 128 \\
\hline
\end{tabular}
\end{table}
% \footnotesize{$^*$ The physical batch size of DP-SGD is set to 128.}\\

To conduct the experiments of Section \textit{DP Methods Comparison}, we pick the best parameters based on the previous sections. The best parameters are those that produce both high accuracy and an ASR as low as possible. Wherever there is a trade-off between accuracy and success rate, we prioritize high accuracy. Table \ref{tab:trainSetup2_mnist} and Table \ref{tab:trainSetup2_cifar} list these parameters for MNIST and CIFAR-10, respectively.

% \FloatBarrier
\begin{table}
\small \caption{Optimal hyperparamters of different DP algorithms for MNIST.  PATE has two different version, based on what parameter used to change the privacy budget of the algorithm; the noises or the number of queries.}
\label{tab:trainSetup2_mnist}
\centering
\small
\begin{tabular}{p{3.5cm} | l }
\hline
 \multicolumn{2}{c}{\textbf{MNIST}} \\
\hline
\textbf{DP-SGD} & Optimizer: SGD (lr=0.1) , Cnorm=2 \\
\hline
\textbf{PATE (Noise-based)} & \#Teachers:200 , \#Queries:10000 , Threshold:150\\
\hline
\textbf{PATE (Query-based)} & \#Teachers:200 , Threshold:150 , Selection noise:120, \\& Result noise:50\\
\hline
\textbf{LP-2ST} & Temperature:0.5, Data split ratio: 50/50\\
\hline
\end{tabular}
\end{table}

% \FloatBarrier
\begin{table}
\small \caption{Optimal hyperparamters of different DP algorithms for CIFAR-10.}
\label{tab:trainSetup2_cifar}
\centering
\footnotesize
\begin{tabular}{p{3.5cm} | l }
% \begin {tabular} {m{3pt} | m{4pt}}
\hline
% \multicolumn{1}{c}{} & 
  \multicolumn{2}{c}{\textbf{CIFAR-10}} \\
%   \hline
% Dataset & \\
\hline
\textbf{DP-SGD} & Optimizer: SGD (lr=0.1) , Cnorm=2 \\
\hline
\textbf{PATE (Noise-based)} & \#Teachers:200 , \#Queries:10000 , Threshold:180\\
\hline
\textbf{PATE (Query-based)} & \#Teachers:200 , Threshold:180 , Selection noise:100, \\ & Result noise:25\\
\hline
\textbf{LP-2ST} & Temperature:0.1, Data split ratio: 40/60\\
\hline
\end{tabular}
\end{table}

% \clearpage
\noindent \textbf{Averaging the Results.}
We discovered that Label-DP algorithms are less stable than the DP algorithms. Therefore, to make the results more unbiased, we repeat their training process 10 times, using different random seeds for the noise, and report the average accuracy and ASR.
The query-based PATE is the PATE model in which noises are constant, and the number of queries is changed to achieve different privacy budgets. In our experiments, we noticed that the results vary among multiple runs. The reason is that the backdoor samples change in different subsets of the queries. Thus in one query subset, the backdoors can be stronger than the other subset. So we repeat the training of query-based PATE 10 times with a random subset of queries selected from 10000 public data points. Finally, we report the average outcome.

\section{Label DP against Backdoors}  \label{sec:labeldp_experiments}
In this section , we evaluate LP-2ST, and ALIBI as two Label-DP models. We investigate if their randomness or other related parameters can help to mitigate the backdoor attacks. To this end, Table~\ref{tab:params_ldp} presents the various parameters involved in these algorithms.
% Label-DP add randomness only to the labels, and keep the feature space intact. Thus it raises this concern that the impact of the Label-DP approaches can be limited on the backdoor attacks in comparison to the regular DP approaches which adding randomness to the input space. We will show that despite this fact, Label-DP methods still can disentangle the association between the trigger patch in backdoor samples and their associated wrong labels, hence mitigate the backdoor attacks. 

\begin{table}
\centering
\caption{\textbf{Parameters of the DP and Label-DP algorithms}}
\label{tab:params_ldp}
\resizebox{\textwidth}{!}{
\begin{tabular}{>{\raggedright\arraybackslash}m{15mm}|m{110mm}}
\hline
\multicolumn{1}{|>{\centering\arraybackslash}m{15mm}|}{\textbf{Method}} 
    & \multicolumn{1}{>{\centering\arraybackslash}m{110mm}|}{\textbf{Parameters}}\\
\hline\hline
   \textbf{LP-2ST}& 
    \begin{enumerate}
        \item \textbf{1. Data split ratio} :  The portion the training dataset split between two training stages (more in the first stage helps with accurate prior but causes underfit in the second stage) 
        \item \textbf{2. Temperature T} :  For logit $z_i$ and calculation of prior $p_i$ of class $i$, a small $T$ in $p_{i}=\frac{exp(z_{i}/T)}{\sum_{j}^{}exp(z_{j}/T)}$ boosts the confidence of the top classes and a large $T$ makes the priors more uniform
        \item \textbf{3. Epsilon $\varepsilon$} : Randomness parameter that is equivalent to the privacy budget 
    \end{enumerate}
\\ \hline
    \textbf{ALIBI} & 
    \begin{enumerate}
        \item \textbf{1. noise of soft training labels} :  Laplacian noise with $\delta=0$ which is applied once and determines the privacy budget
    \end{enumerate}
\\ \hline
\end{tabular}}
\end{table}

\subsection{LP-2ST vs. Backdoors}
Since the Label-DP algorithms randomly change the labels, we found that the accuracy in high noise fluctuate among multiple runs. So for each experiment on LP-2ST and ALIBI, the accuracy and ASR are the averages of 10 trials. For each figure from left to right, we pick a parameter bold on the x-axis (which are chosen randomly) and apply it for the experiments in the succeeding figure. For the first two figures, we set $\varepsilon=1$.

\begin{figure*}
  \centering
    \includegraphics[width=\textwidth]{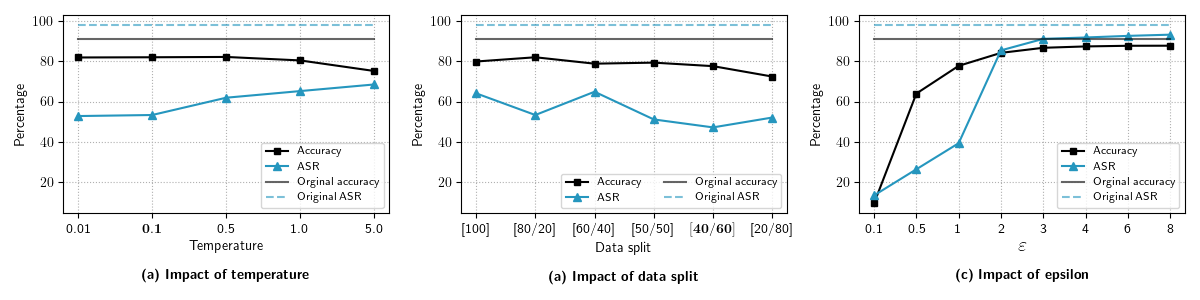}
        % \vspace{-1.5em}
    \caption{\textbf{The impact of temperature, data split between two stages and epsilon on LP-2ST (from left to right).} Epsilon, the factor of privacy-preserving in LP-2ST, can drastically deteriorate the ASR with an acceptable utility cost (c).
    }
    \label{fig:MLDP}
        % \vspace{-2em}
\end{figure*}

Fig. \ref{fig:MLDP}a demonstrates the effect of temperature with a random data split of [80/20]. Compatible to \cite{ghazi2021deep}, sparsifying the priors helps to improve the utility, but to our surprise, it decreases ASR. We speculate the reason is that the backdoor still has a touch of the base class. Thus the first round of LP-2ST predicts target and base classes as the backdoors' top-2 classes. The sparsified prior shifts the probabilities of these two classes far away from zero, so the algorithm selects the base class more confidently.

In Fig. \ref{fig:MLDP}b the training data has been partitioned for two stages. [p1/p2] on the x-axis indicates the percentage of the data in stage 1 and stage 2 of LP-2ST, respectively. When 100\% of data is allocated to the first stage, it means that we are using LP-1ST with RR. There is not a clear pattern between ASR and data split. But an LP-2ST model with more data in the first stage has more enhanced priors and higher accuracy.

Fig. \ref{fig:MLDP}c compares different privacy budgets $\varepsilon$, which is the random factor of the RR algorithm. Naturally, more randomness helps to decrease the ASR. Especially the results for $\varepsilon=1$ are impressive since it drops the ASR to less than 40\%, while the accuracy is still roughly 80\%.

\textbf{Conclusion:} To our surprise, eventhough Label-DP only randomize the labels, but it is still successful against backdoor attacks. In this success all parameters are involved but noise has the major impact. LP-2ST vividly can mitigate the attack but it is very important what $\varepsilon$ is selected to obtain a reasonable accuracy-ASR trade-off.

\subsection{ALIBI vs. Backdoors}
\begin{figure}
  \centering
    \includegraphics[width=60mm]{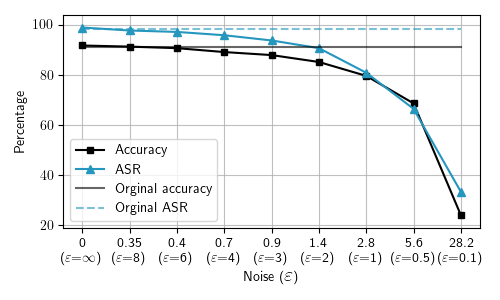}
    % \includegraphics[width=50mm]{figures/ALIBI_noise_2.png}
    % \includegraphics[width=50mm,scale=0.5]{method.eps}
        % \vspace{-1.5em}
    \caption{\textbf{Effectiveness of randomizing labels on reducing ASR in ALIBI}. The noise added to one-hot labels in ALIBI impacts both accuracy and ASR proportionally.
    }
        % \vspace{-1.75em}
    \label{fig:ALIBI}
\end{figure}

According to Fig. \ref{fig:ALIBI}, ALIBI with higher noise drops both accuracy and ASR proportionally. It can be justified by the fact that all the labels randomly change just once at the beginning of the training.

\textbf{Conclusion:} On average, ALIBI can mitigate the effect of backdoor attacks but with reduced utility costs.

\subsection{Training Process}

In this section, we compare the training process of DP-SGD, LP-2ST, and ALIBI on CIFAR-10. These comparisons are based on two privacy budgets $\varepsilon=\infty$ and $\varepsilon=1$, to provide an overview over the training process with and without randomness. For LP-2ST, we only illustrate the training of the second and final stage of the algorithm. 

\begin{figure*}
  \centering
     \includegraphics[width=\textwidth]{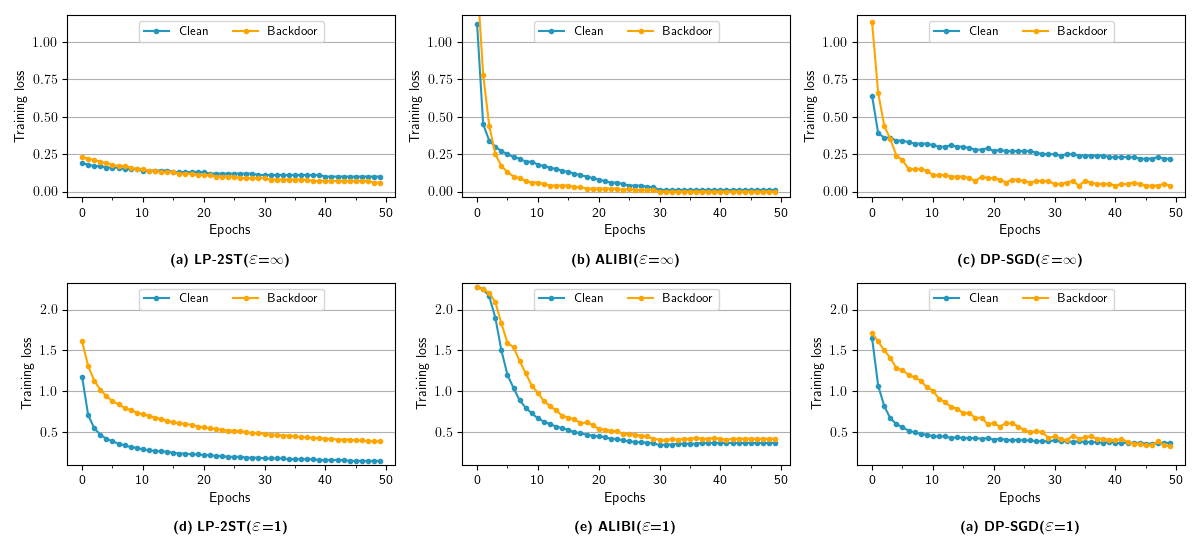}
        %  \vspace{-1.75em}
    \caption{\textbf{An overview of the training process of LP-2ST, ALIBI and DP-SGD using $\varepsilon=\infty$ (upper) and $\varepsilon=1$ (lower).}
    }
    \label{fig:training}
        % \vspace{-1.75em}
\end{figure*}

In Fig. \ref{fig:training}, each column demonstrates a different method, and each row indicates one of the privacy budgets. For all three differentially private methods, on the first row, with $\varepsilon=\infty$, the loss of the backdoor samples drops below the clean loss on early training epochs. It is the opposite for all three methods when $\varepsilon=1$ on the second row. For LP-2ST the backdoor loss does not converge to the clean loss and remains higher. 
It is consistent with the results of LP-2ST at $\varepsilon=1$ in Fig. \ref{fig:MLDP}c. 
For ALIBI the clean and backdoor losses are changing very closely. It explains the similar values for the ALIBI accuracy and ASR in Fig. \ref{fig:ALIBI}. DP-SGD can resist the backdoor samples on early epochs. So one of the suggestions is to stop the training early to avoid backdoors to overfit.

\textbf{Conclusion:} During DP training, the model underfits or suppresses the backdoor samples which results in defusing the backdoors' impact on the model. This finding confirms the results of the paper.
% The comparison of different methods illustrates the resillience that each DP method provide against backdoor examples during training process which . 

\end{document}